# The dark and dusty side of galaxy evolution

Stephen Serjeant

*Department of Physics and Astronomy, The Open University, Milton Keynes, MK7 6AA, UK*

**Abstract.**
In this brief, partial, incomplete and egregiously self-citing review I will summarise some of the key results in the past few years in surveys for dusty star-forming galaxies and some exciting prospects for forthcoming surveys.

## INTRODUCTION

The purpose of this presentation is to give some background and historical context for the subsequent presentations in this conference. There is far too much material to do justice to the whole field, so this review will necessarily be incomplete and partial, but I'll cover a few key results with particular contemporary relevance. A recurrent theme is resolving the populations that dominate the observed extragalactic background light. You might reasonably ask why we should care about this apparently observer-dependent population. The extragalactic background light $I_\nu$ is an integral over the comoving luminosity density $\varepsilon$: $I_\nu(\nu_0) = \frac{1}{4\pi} \int \frac{\varepsilon((1+z)\nu_0, z)}{1+z} dr_{\mathrm{comoving}}$. Therefore, at any fixed redshift, the galaxies which dominate the background also dominated the luminosity density. The extragalactic background light is therefore closely linked to the comoving volume-averaged star formation history.

## IRAS IN TWO MINUTES

Where do you start with a mission that has generated about 180,000 citations? For me some of the most important discoveries have been the radio-far-infrared correlation of star-forming galaxies [1] and the discovery of ultraluminous galaxies (ULIRGs, $10^{12-13} L_\odot$). Highly disturbed optical morphologies correlate with high star formation rate, and vice versa (e.g. [2]). It's often argued that local ULIRGs are still useful to study as templates for high-redshift objects, but I believe this misses the point: what's interesting is how they differ from high-redshift objects.

The local $60\,\mu$m luminosity function [3] is also a key constraint of galaxy evolution models, though direct constraints at other far-infrared wavelengths are more sketchy. However it's still possible to use the submm/far-IR colours of galaxies together with the observed IRAS colours to predict the submm luminosities of the IRAS PSCz survey and derive luminosity functions [4, 5] that will shortly be tested with Herschel.

Another surprise from IRAS was the discovery of so-called Hyper-luminous[1] infrared galaxies (HLIRGs) with $> 10^{13} L_\odot$. HLIRGs were originally posited as examples of monolithic collapse, which at the time was considered an alternative to heirarchical formation; to this day the semi-analytic community resist calling downsizing "anti-heirarchical". The prototype HLIRG was IRAS FSC 10214+4724 [6]. Later investigation showed this galaxy is gravitationally lensed [7, 8, 9], raising the question as to whether all HLIRGs are lensed, which turned out not to be the case [10]. Nevertheless Herschel may find many lensed HLIRGs as we'll see.

## ISO IN ONE MINUTE

ISO had many breakthroughs (see [11] for a review), most notably in my view the opening up of the mid-infrared spectroscopic window for diagnosing the relative bolometric contributions of star formation and active nuclei in nearby galaxies (e.g. [12]) which we will see much of at higher redshift in this conference, the discovery of a strongly evolving mid-infrared-selected population of star-forming and active galaxies (e.g. [13, 14, 15, 16]), and memorable demonstrations of spatially-resolved highly obscured star formation in nearby galaxies (e.g. [17]).

## SCUBA AND SUCCESSORS IN TWO MINUTES

The surprising discovery of mJy-level galaxies in the first submm galaxies was one of the first indications of downsizing (e.g. [18, 19, 20]) and either requires radical changes to the IMF or to models of feedback processes. This population was not predicted by prior theoretical models. Source counts in blank-field surveys to ultra-deep mapping of gravitational lens clusters can account for all the $850\,\mu$m background (e.g. [21]) though there are very few resolved galaxies at the faint end; stacking analyses intruigingly account most of the $450\,\mu$m background but less than half that at $850\,\mu$m (e.g. [22]). In the interregnum between SCUBA and SCUBA-2, an impressive LABOCA $870\,\mu$m analysis [23] found the first environmental dependence of submm star formation, and the first blind CO redshift of a submm galaxy [24] is an exciting taste of the future.

## SPITZER AND AKARI IN TWO MINUTES

Spitzer has pushed the spectroscopic mid-infrared diagnostics to higher redshifts (e.g. [25]) demonstrating that submm galaxies' bolometric luminosities are dominated by star formation rather than black hole accretion. Spitzer has also been the source of new unexpected populations (e.g. [26, 25]) and has clearly demonstrated downsizing in star formation (see [27, 28]). AKARI only had a 2-year cooled mission compared to Spitzer's five, with the time dominated by its all-sky survey, but it nevertheless conducted a small but impressive suite of deep surveys. The filter set spans Spitzer's $8-24\,\mu$m gap making photo-$z$ possible from redshifted PAH features [29]. Ultra-deep mapping of a galaxy cluster gravitational lens recently resolved most of the expected $15\,\mu$m background [30].

---

[1] As a joke I once tried to coin the phrase Über-luminous for $> 10^{14} L_\odot$ galaxies, but the referee would have none of it, perhaps quite rightly.

## BLAST IN ONE MINUTE

BLAST's data is exquisitely beautiful though I don't like some of the things they've done with it. For example, the claim has been made that the BLAST surveys at $250, 350, 500\,\mu$m have accounted for all the submm extragalactic background through stacking analyses [31, 32]. This assumed the Spitzer-detected populations they're stacking are unclustered, but in my view the evidence presented for this lack of clustering [32] is unconvincing (see [33] and section 36 of [34]). Following the method of [22], an integral of the two-point angular correlation function $w(\theta)$ can yield an estimate of the overestimate made from stacking a clustered population: $S_\nu = I_\nu \int_0^\infty w(\theta)B(\theta)2\pi\theta d\theta$ where $B$ is the telescope beam. Even the most conservative assumptions about the clustering of the Spitzer population yield a factor of two overestimate in the stacked backgrounds of [32]. However other parts of the BLAST analysis are very interesting, such as the demonstration from clustering measurements that star formation at $z > 1$ is preferentially located in the outskirts of groups and clusters [35].

## HERSCHEL IN TWO MINUTES

The successful launch of the ESA Herschel mission in May 2009 promises a tremendous step forward in survey capabilities. At the time of writing there is little that can be presented in public, though the thrilling first images from the SPIRE instrument at $250\,\mu$m of local galaxies already show the background sky is very quickly confusion limited. The HerMES and PEP GTO surveys are making a carefully constructed "wedding cake" of tiered wide/shallow and narrow/deep surveys to maximise the coverage of the submm luminosity – redshift plane. HerMES will clearly need to make use of image reconstruction techniques at the longer wavelengths of all but the widest tiers. In open time the largest survey is the $550\,\text{deg}^2$ Herschel ATLAS, sometimes called the "SLOAN of the submm", which will dominate both the total number of sources and sky area mapped for extragalactic sources by Herschel. The steep bright source count slopes in the submm imply a strong gravitational lensing magnification bias; the Negrello et al. 2007 models predict several hundred strong lenses in H-ATLAS that can be identified with an astonishing $\sim 100\%$ efficiency (c.f. 0.1% in CLASS). These will be obvious follow-up targets for eMERLIN and submm/mm-wave interferometry. H-ATLAS will measure the dust mass function of local galaxies as low as $10^{4.5}\text{M}_\odot$ and will measure or constrain the star formation rates of thousands of quasars. The indications from stacking analyses and a few direct detections are that quasar host galaxies have a comoving volume-averaged star formation history evolving as fast or faster than ULIRGs, as $(1+z)^8$ at $-26 > I_{\text{AB}} > -28$ out to $z \sim 1.5$ (adapting data from [36]).

## SCUBA-2 IN ONE MINUTE

Like Herschel, SCUBA-2 promises an imminent revolution in submm-wave survey capabilities. The submm galaxy clustering measurements from the SCUBA-2 Cosmology Survey should easily achieve what the SHADES survey attempted: to test whether submm galaxies are consistent with being the progenitors of present-day giant ellipticals. The ultra-deep $450\,\mu$m survey from this and other surveys will determine whether mid-infrared populations do indeed dominate the $450\,\mu$m background as stacking analyses suggest (e.g. [22]). The SCUBA-2 All-Sky Survey, meanwhile, is searching for lenses and new populations: is there a limit to the integrated star formation rate of a galaxy? Are there new populations of more massive galaxies forming the bulk of their stars at $z > 4$, or do all massive galaxies obey the putative peak in comoving star formation rate at $z \sim 2$?